\documentclass[12pt,a4paper]{article}

\usepackage[a4paper]{geometry}
\newcommand{\beq}{\begin{equation}}
\newcommand{\eeq}{\end{equation}}
\usepackage{amsmath,graphicx,amssymb}

\title{The effect of Kolmogorov (1962) scaling on the universality of
turbulence energy spectra} 
\author{W. D. McComb and M. Q. May,\\
SUPA School of Physics and Astronomy,\\
Peter Guthrie Tait Road,\\
University of Edinburgh,\\
EDINBURGH EH9 3JZ.\\
Email: wdm@ph.ed.ac.uk}

\begin{document}
\maketitle

\begin{abstract}

It has long been established that turbulence energy spectra scale on the
Kolmogorov (1941) variables over a wide range of Reynolds numbers and in
vastly different physical systems,  depending only on the  dissipation
rate, the kinematic viscosity and the wavenumber. On the other hand, the
analogous study of structure functions in real space is strongly
influenced by the Kolmogorov (1962) \emph{refined} theory, which
introduced a dependence on a large length scale $L_{ext}$,
characteristic of the system size. If such a dependence exists it is
surprising that it does not show up in the study of wavenumber spectra,
where the different physical systems suggest that $L_{ext}$ can vary by
up to five orders of magnitude. Here we use an order of magnitude
calculation to suggest that scaling according to Kolmogorov (1962) would
destroy the observed asymptotic universality of energy spectra at large
wavenumbers.

\end{abstract}

\thispagestyle{empty}
\setcounter{page}{0}
\newpage

\section{Introduction}

There is a curious dichotomy at the heart of turbulence theory between
real-space and wavenumber-space formulations, one being merely the
Fourier transform of the other. It has its origins in the work of
Kolmogorov, who in 1941 obtained expressions for the second- and
third-order structure functions \cite{Kolmogorov41a,Kolmogorov41b}:
referred to as K41. In particular, he used dimensional analysis in
combination with Richardson's picture of a cascade of eddies, to obtain
an expression for the second-order structure function of the form:
\beq
S_2(r)=C \varepsilon^{2/3} r^{2/3}.
\label{Kol_41}
\eeq
This applied to a range of length scales $r$ intermediate between the
large scales associated with the size of the physical system and  the
small scales determined by the dissipative action of viscosity. The
prefactor $C$ is taken to be constant and $\varepsilon$ is the mean energy
dissipation rate.

In 1945, Onsager \cite{Onsager45} applied similar arguments in
wavenumber space to obtain
the corresponding energy spectrum in the form:
\beq
E(k)  = \alpha \varepsilon^{2/3} k^{-5/3},
\label{spect41}
\eeq 
where $\alpha$ is constant and $k$ is the wavenumber. This is the
legendary \emph{Kolmogorov -5/3 law}, while $\alpha$ is the
\emph{Kolmogorov constant}. This result can also be obtained by Fourier
transformation of (\ref{Kol_41}).

\begin{figure}  \begin{center}
\includegraphics[width=1.0\textwidth]{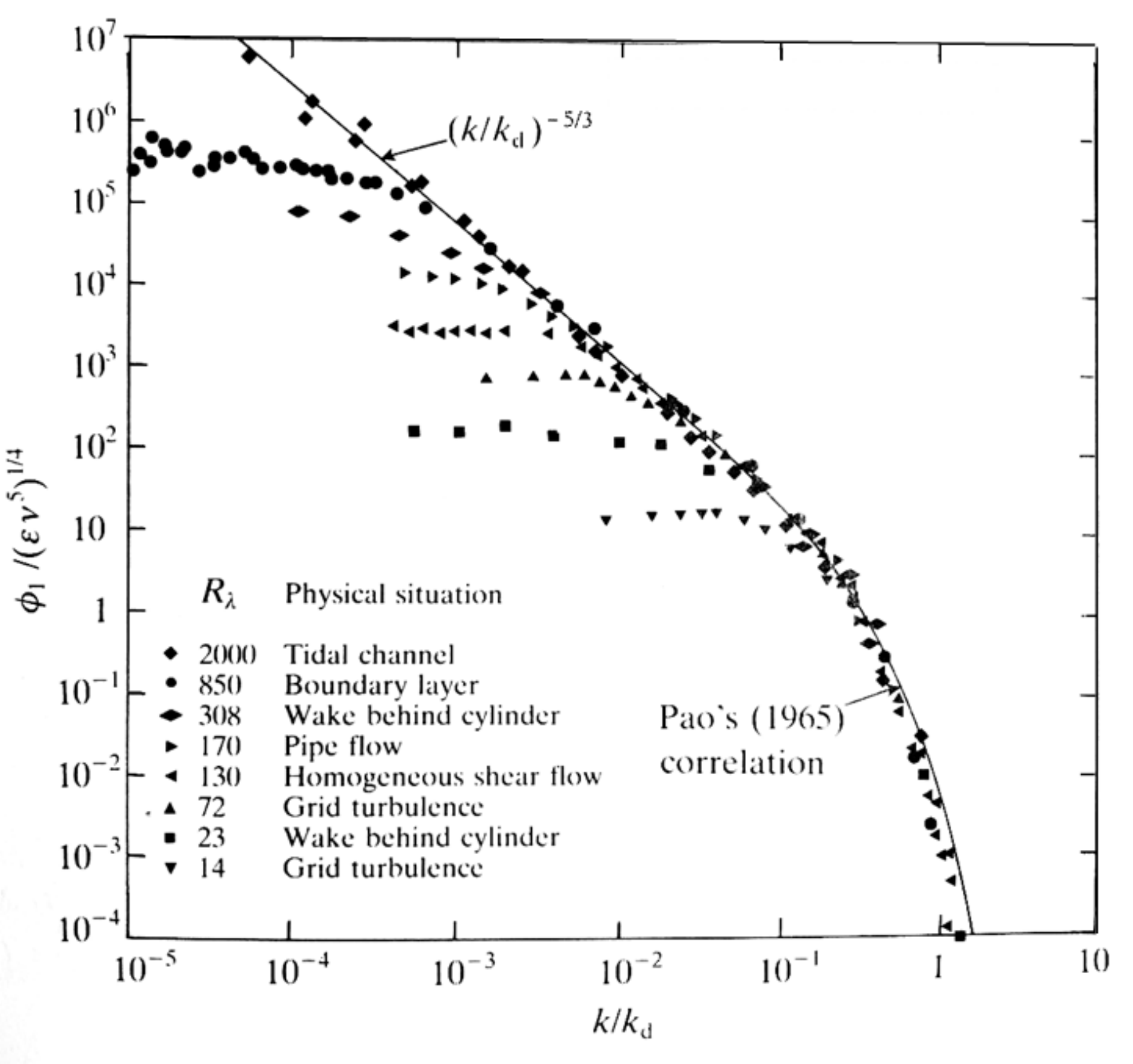} \end{center} 
\caption{\small  Measured one-dimensional energy spectra for a wide
range of Reynolds numbers and physical situations showing the asymptotic
effect of scaling on Kolmogorov (1941) variables. Image reproduced from
Figure 2.4 of Reference \cite{McComb90a}.}  \label{PFTspec} 
\end{figure}

The subsequent history of the subject is of interest. The work of
Kolmogorov, originally only known in Russia, was brought to the wider
fluid dynamics community by Batchelor, beginning with a paper discussing
the real-space treatment of the subject in 1947 \cite{Batchelor47}.
However, after this Batchelor worked in wavenumber $(k)$ space,
particularly in his classic monograph \cite{Batchelor71}, and this set a
pattern for later researchers. The situation may be summarised by Figure
2.4 of Reference \cite{McComb90a}, which is reproduced here
for the reader's convenience as Fig. \ref{PFTspec}. What is actually shown here is the
one-dimensional projection of the spectrum, $\phi_1(k)$, which has been made
dimensionless using the Kolmogorov variables and plotted against
wavenumber divided by the Kolmogorov dissipation wavenumber $(k_d)$. The
various flows are characterised by their value of the Taylor-Reynolds
number $R_\lambda$, with values shown over a range from 14 to 2000. The
existence of universal asymptotic behaviour at high wavenumbers is
clearly established by this set of graphs, as is the -5/3 law. In
particular, the results of Grant \emph{et al.} in 1962, showing the -5/3
law over many decades of log wavenumber, clearly established the
correctness of the Kolmogorov theory. Paradoxically, in the same year
Kolmogorov decided that his theory was wrong \cite{Kolmogorov62}.

\section{Kolmogorov's refined (\emph{sic}) theory of 1962}

Kolmogorov's original theory \cite{Kolmogorov41a,Kolmogorov41b} was
based on the so-called \emph{Richardson picture}, in which the
turbulence energy transfer proceeded by a step-wise cascade from large
eddies to small. Bearing in mind that this is a random process, it could
be argued that the average effect would be that detailed information
about the  large scales would be lost. Accordingly the behaviour of the
small scales would be universal. However, Kolmogorov (following a rather
obscure criticism of K41 by Landau) introduced a length scale $L_{ext}$
said to be characteristic of the size of the system and modified the
result (\ref{Kol_41}) for the structure function to
\beq
S_2(r)=C \varepsilon^{2/3} r^{2/3}(L_{ext}/r)^{-\mu}.
\label{Kol_62}
\eeq

Accordingly the title of the later work \cite{Kolmogorov62}, which refers to
`A refinement of previous hypotheses \dots', is something of a misnomer.
In fact the basic idea of independence from initial conditions has been
abandoned. Evidently the presence of $L_{ext}$ in the result destroys
any claim to universality.

As equation (2) may be derived from (1) by Fourier transformation, so may we
derive the K62 version of the energy spectrum the same way by
transforming equation (3) to obtain
\beq
E(k)  = \alpha \varepsilon^{2/3} k^{-5/3}(L_{ext}k)^{-\mu}.
\label{spect62}
\eeq
It is intuitively obvious that this form, if correct, could pose
problems for the spectral scaling shown in Figure 1. Also, we should not
forget that K41 is an asymptotic result which applies also to the
dissipation region, where the spectrum (irrespective of its detailed
form) must scale as a function $f(k/k_d)$.
The main objective
of this paper is to make a quantitative test of this, but first we will
briefly review the development of the subject subsequent to 1962. In
particular, we shall discuss \emph{intermittency corrections} and
\emph{anomalous exponents}.

\section{Anomalous exponents}

Although the original K41 theory was restricted to the
second- and third-order structure functions, the various arguments and
hypotheses were applied to the probability distribution, and so the
theory is as applicable to structure functions of all orders. Hence we
now introduce the general form of the longitudinal structure functions as:
\beq
S_n(r) = \left\langle \delta u^n_L(r) \right \rangle,
\eeq
where the longitudinal velocity increment is given by
\beq
\delta u_L(r) = [\mathbf{u}(\mathbf{x} + \mathbf{r},t) -
\mathbf{u}(\mathbf{x},t)]\cdot \mathbf{\hat{r}}.
\eeq
If for $n \geq 4$ these structure functions are found to exhibit power
laws, then dimensional analysis would lead to
\beq
S_n(r) \sim (\varepsilon r)^{n/3}.
\eeq
However, measurement of the structure functions has repeatedly found a
deviation from the above dimensional prediction for the exponents. If
instead, the structure functions are taken to scale with exponent
$\zeta_n$, thus
\beq
S_n(r) \sim r^{\zeta_n},
\eeq
then it has been found  \cite{Anselmet84,Benzi95} that the difference
$\Delta_n = |n/3 - \zeta_n|$ is nonzero and increases with order $n$. In
particular, Figure 14 of Reference \cite{Anselmet84} illustrates this
behaviour rather nicely. 

At first this behaviour was classed as \emph{intermittency corrections}
and the concept of intermittency was associated with the very small
scales where the dissipation was mainly concentrated. Later it was
recognised that intermittency was present at all scales and nowadays the
tendency is to speak of \emph{internal} rather than \emph{small-scale}
intermittency. In any case, since the mid-1990s, it has become usual to
refer to the $\zeta_n$ as anomalous exponents, and they are seen as a
subject of fundamental interest to this day (e.g.
\cite{Mailybaev12},\cite{Saw18}). 

\section{Testing the effect of K62 on spectral scaling}

\begin{figure} 
\begin{center}
\includegraphics[width=1.2\textwidth]{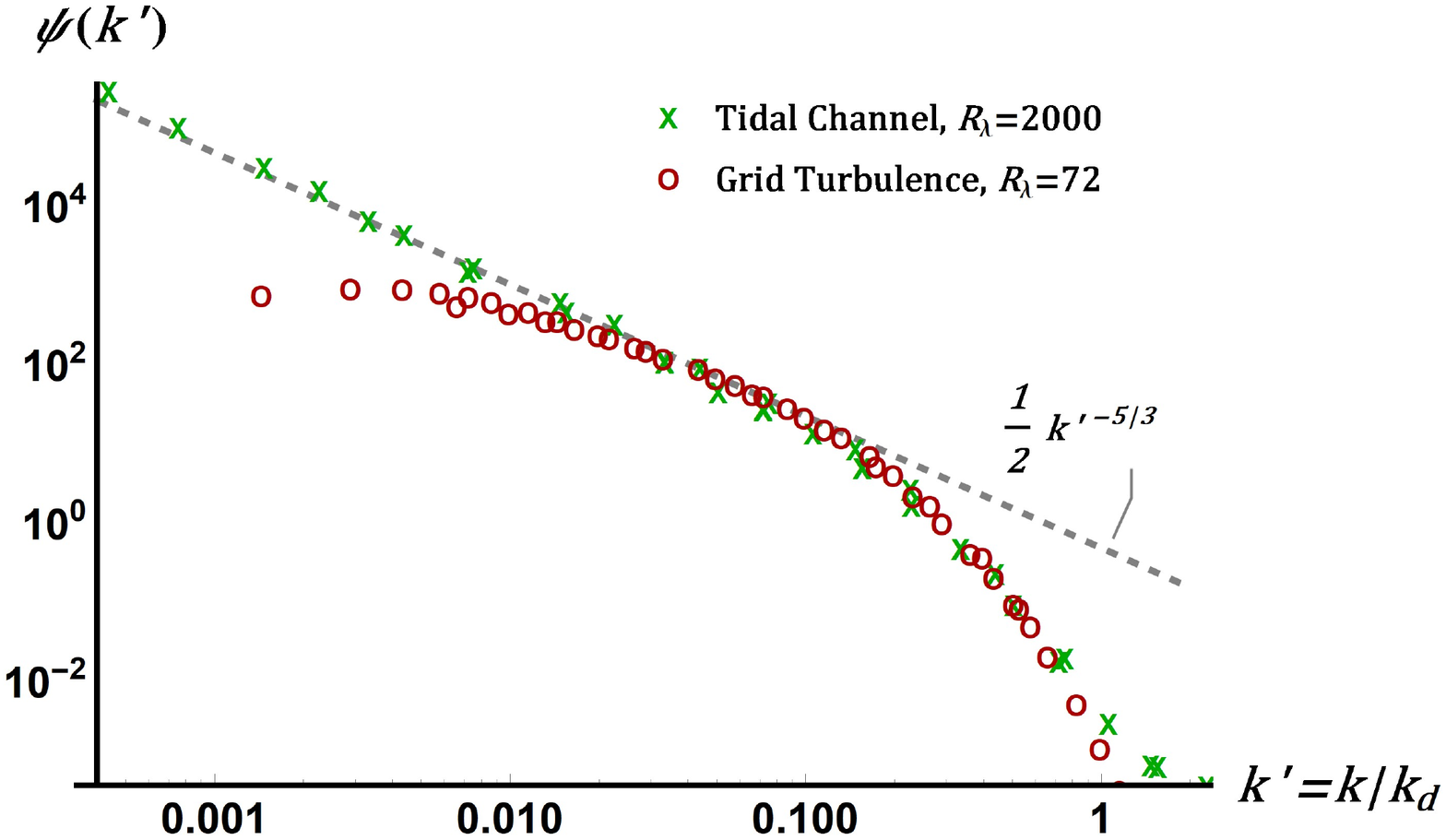}
\end{center} 
\caption{\small Two representative measured one-dimensional energy
spectra, one at $R_\lambda = 2000$, taken in a tidal channel
\cite{Grant62}; the other at $R_\lambda = 72$, taken in grid generated
turbulence \cite{Comte-Bellot71}; and normalised on Kolmogorov (1941) variables. Note that
we plot the \emph{scaled} one-dimensional
spectrum, as given by $\psi(k')=\phi(k')/(\varepsilon \nu^5)^{1/4}$.} 
\label{genspec} 
\end{figure}

\begin{figure} 
\begin{center}
\includegraphics[width=1.2\textwidth]{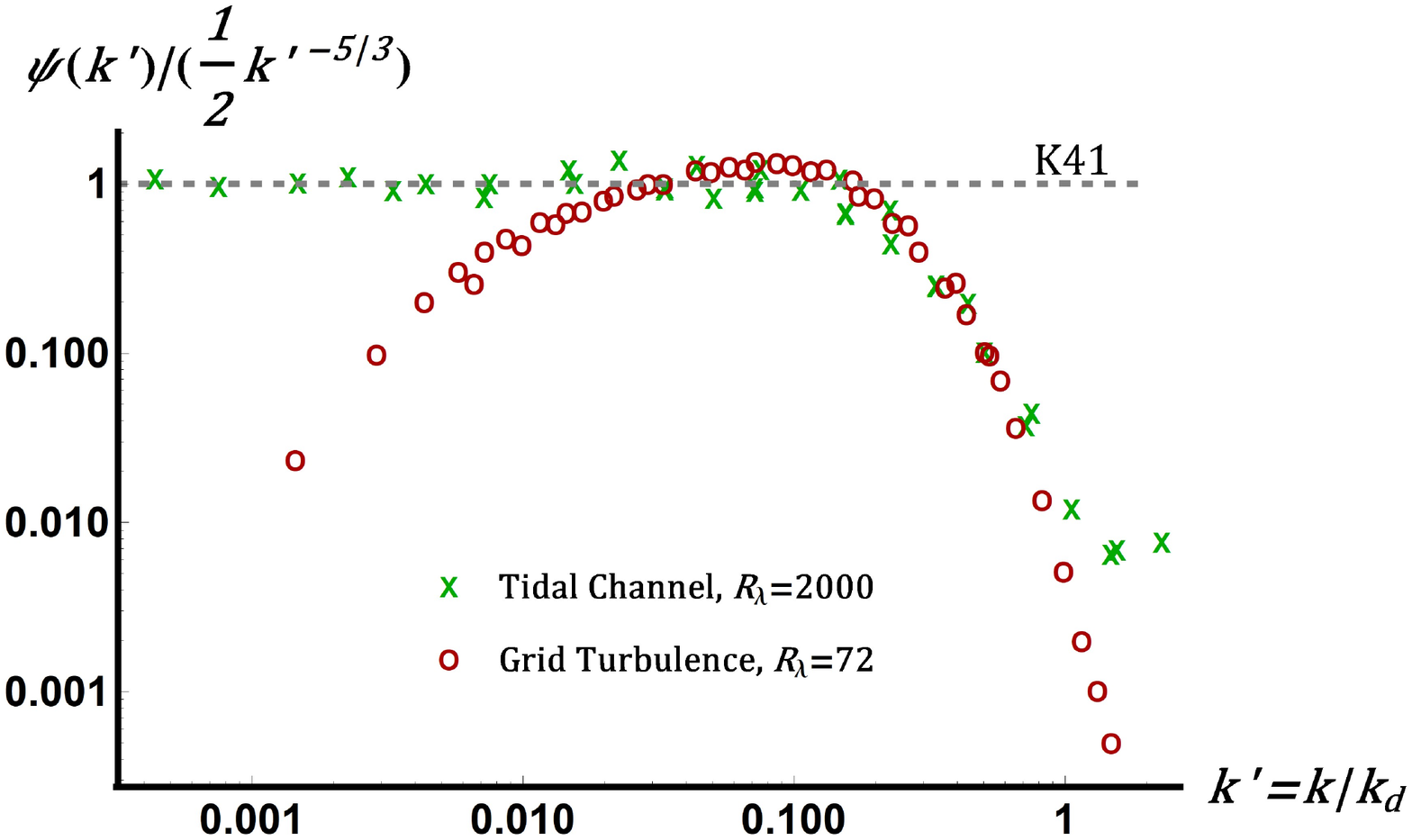}
\end{center} 
\caption{\small The results of Figure \ref{genspec} plotted in
compensated form 
such that $k^{-5/3}$ spectra appear as a constant. Note that we have
taken the one-dimensional spectral constant to be $\alpha_1 = 1/2$, on
the basis of Figure \ref{genspec}.} 
\label{compspec} 
\end{figure}

In order to assess the effect of changing from K41, as given by
(\ref{Kol_41}), to K62, as given by (\ref{Kol_62}), we need to make
estimates of $L_{ext}$ and $\mu$. We begin by choosing two disparate
investigations in order to make a comparison. We choose the results of
Grant \emph{et al.} \cite{Grant62}, which were taken in a tidal channel
at a Taylor-Reynolds number of $R_{\lambda} = 2000$, and the laboratory
results of Comte-Bellot and Corrsin \cite{Comte-Bellot71} taken in grid
turbulence (with two-inch grid) at $R_{\lambda}= 72$. We plot these
results on log scales in Figure \ref{genspec} as the \emph{scaled}
one-dimensional spectrum $\psi(k')=\phi(k')/(\varepsilon\nu^5)$ against
the dimensionless wavenumber $k'=k/k_d$, where $k_d$ is the Kolmogorov
dissipation wavenumber. Here we have taken the one-dimensional
Kolmogorov constant as $\alpha_1 = 1/2$, as this gives good agreement
with both sets of results. This is shown in Figure \ref{compspec},
where we follow the modern practice of plotting the spectrum divided by
the Kolmogorov form, such that K41 corresponds to a horizontal line at
unity. 

In the absence of a precise definition of $L_{ext}$, we can determine
$L'_{ext} = 2\pi / k'_{ext}$, where $k'_{ext}$ marks the departure of
the curve from the K41 form, as one goes from high wavenumbers to low.
In this case we estimate $L'_{ext} \sim 50$ for the grid turbulence, and
$L'_{ext} \sim 2000$ for the tidal channel measurements. In fact, the
spectra in the results of Grant \emph{et al.} \cite{Grant62} do not
actually peel off from the $-5/3$ line at low $k$ and so our estimate is
actually a lower bound for $L_{ext}$ in this case, and this favours K62
in the comparision. 

The exponent $\mu$ is seen as a universal feature of the K62 theory and
gnerally estimates have been in the range $0.1 - 0.2$. We have taken
the value $\mu=0.1$ as reported by Kaneda \emph{et al.} \cite{Kaneda03}
from their high-resolution numerical simulation. Again, this use of the
lower value favours K62 in the comparison.

\begin{figure} 
\begin{center}
\includegraphics[width=1.0\textwidth]{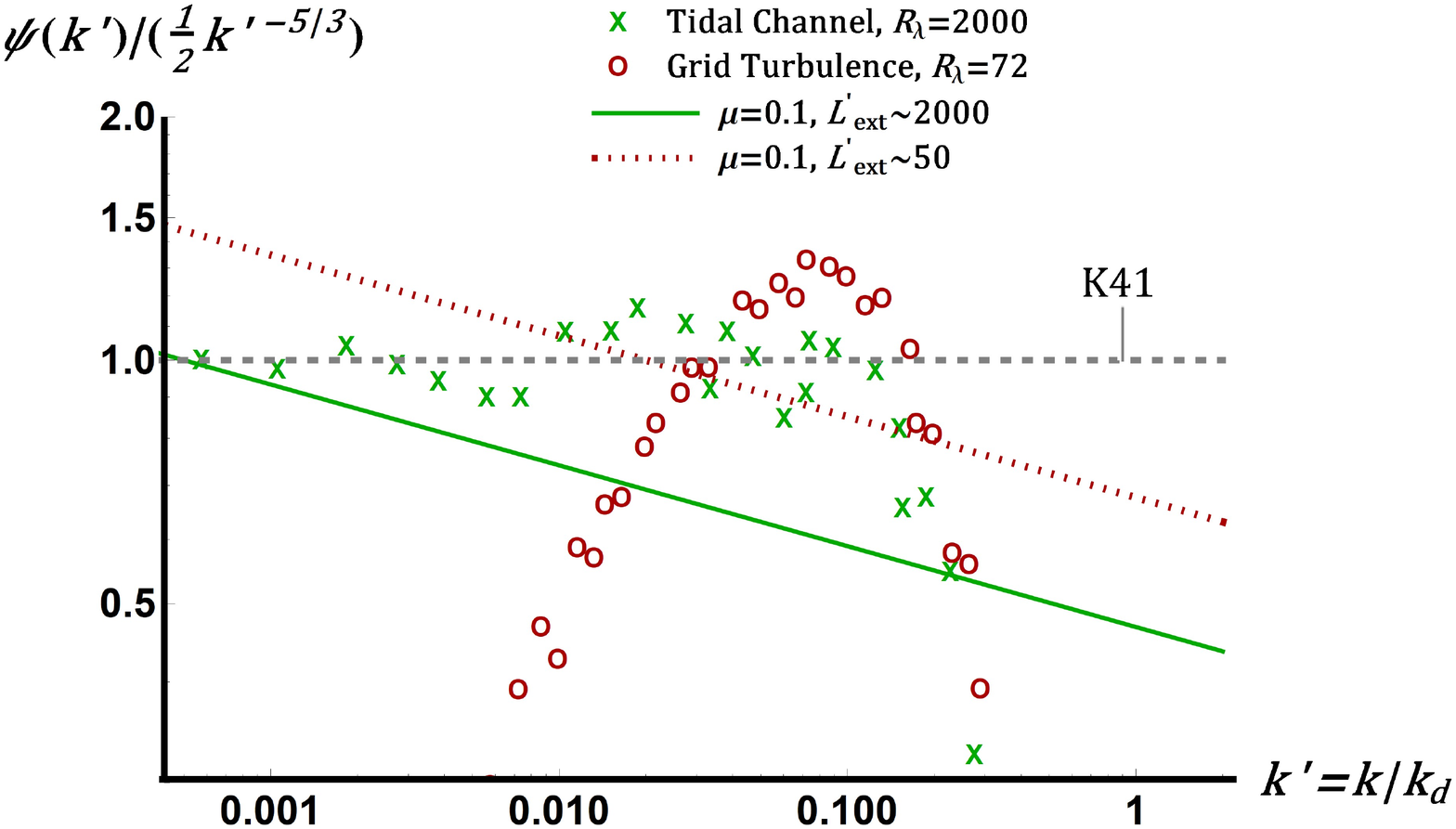}
\end{center} 
\caption{\small Demonstration of the effect on the compensated spectra
of Figure \ref{compspec} of including the Kolmogorov (1962)
corrections to the K41 theory. The departure of the slope from K41 was
based on the value of $\mu$ obtained in the numerical simulation of
Kaneda \emph{et al.} \cite{Kaneda03}.} 
\label{K62spec} 
\end{figure}

The result of changing from (\ref{spect41}) to (\ref{spect62}) can be
seen in Figure \ref{K62spec}. Plotting a compensated spectrum, based on
Kolmogorov variables, for a form like equation (\ref{spect62}) evidently
leaves a residual slope, given in this case by $\mu=0.1$. For the values
of $L_{ext}$ taken here, the two scaled spectra no longer coincide but
indeed differ in a constant ratio of $0.69$. It should be noted that
the change of ordinate scales in this plot exaggerates the spread of
values about the line corresponding to K41, and also emphasises the bump
at $k'=0.1$ which is characteristic of spectra at low Reynolds number
and which disappears with increasing Reynolds number (e.g. see Reference
\cite{Ishihara09}).

\section{Discussion}

We should point out that the fact that Kaneda \emph{et al.}
\cite{Kaneda03} measured a slope different from $-5/3$ is not evidence in
favour of K62. Such measurements, with a finite inertial range, are
sensitive to the criteria used to establish the extent of that range. In
practice, it is expected that at large wavenumbers spectra will roll off
in some form of exponential. It is also worth pointing out that, as a
modification to the power law requires the introduction of a length
scale in order to preserve the correct dimension, internal scales like
the Taylor microscale or the Kolmogorov dissipation length scale, if used
for this purpose, will also change the dependence on the dissipation
rate.

It seems surprising that there is, and has been, such a large
concentration on intermittency, when K41 relies on various
approximations, in particular the neglect of the viscosity. In 2002,
Lundgren offered a rigorous proof that K41 was valid in the limit of
infinite Reynolds number \cite{Lundgren02}. This was reinforced by the
author's comparison with the experimental results of Mydlarski and
Warhaft \cite{Mydlarski96}. More recently various workers have given
some attention to the effects of finite viscosity and finite system size
\cite{Antonia06}--\nocite{Tchoufag12}\nocite{McComb14b}\nocite{Meldi17}\cite{Antonia17}
and of the neglect of the time dependence \cite{McComb18a} in K41.  
 
Many of the problems in this topic seem to arise from the addiction of
the turbulence community to real-space treatments\footnote{Of course the
vast majority of applications in engineering fluid dynamics require
real-space treatments, but we are concerned here with fundamentals.}. If
the cascade is seen as unsatisfactory then recourse is made to  a vortex
stretching picture: see Reference \cite{Antonia82} for a general
discussion. However, although both pictures have their own intuitive
appeal, they both suffer from a certain vagueness. Indeed their general
`hand waving' characteristics do not provide a satisfactory basis for a
mathematical physics approach.

The contrast with the wavenumber picture could not be greater. Here the
Fourier modes are the independently excited degrees of freedom of the
system. The number of these degrees of freedom increases with the
Reynolds number and, in the limit of infinite Reynolds number, there is
an infinite number of them and (naturally!) an infinite
amount of energy. They are all coupled to each other by the nonlinear
term (nonlinear mixing) and, in the language of statistical physics,
constitute a \emph{many-body problem}. The energy-conservation of the
nonlinear term can be deduced by inspection, and the existence of the inertial
range can be deduced by simple physical reasoning. It corresponds to the
case where the \emph{energy flux} through wavenumber attains a constant,
maximum value, which in a steady state is equal to the viscous
dissipation rate. All of this can be found in the monograph by Batchelor
\cite{Batchelor71} which was originally published in 1954. It also tells
us that the Kolmogorov $-5/3$ spectrum is due to the scale invariance of
the energy flux and, as this is a globally averaged quantity, it is
unaffected by intermittency, but rather takes it into account. To
conclude this point, all fundamental statistical closures of turbulence
are formulated in wavenumber space, all are dimensionally compatible
with K41, and differences between them boil down to the existence or
otherwise of certain integrals.

\section{Conclusion}

The behaviour of the exponents of the structure functions, as shown in
Figure 14 of Reference \cite{Anselmet84}, is often cited as evidence for
anomalous exponents, and possibly for K62. However, alternative
explanations are the increasing sensitivity of higher-order moments to
the approximations made in K41 or to systematic error. In the latter
case, given a fixed approximation to the probability distribution,
evaluating exponents of progressively higher order should indeed yield
progressively larger deviations from the canonical results. This
possibility has been tested by McComb \emph{et al.} \cite{McComb14b},
for the particular case of the second-order structure function, by
making allowance for systematic error. In this way they found that the
second-order exponent tended to the K41 form, in the limit of infinite
Reynolds numbers.

Taken in conjunction with the results presented in Fig. \ref{K62spec},
this suggests a more cautious approach to the whole idea of anomalous
exponents. It is our view that the papers published in 1941 represent
the true legacy of A. N. Kolmogorov, and that the 1962 `refinement' was
unnecessary.

\section{Acknowledgements}

We would like to thank Bob Antonia, Jorgen Frederiksen, Marcello Meldi
and Sam Yoffe for reading a first draft of this work and for making
helpful comments and suggestions.



\end{document}